 \documentclass[final,5p,times,twocolumn]{elsarticle}

\usepackage{amssymb}
\usepackage{lipsum}

\usepackage{hyperref}
\usepackage{graphicx}
\usepackage{amsmath}
\usepackage{amsfonts}

\journal{Physics Letters B}

\newcommand{\GeV}{\makebox{ GeV}}
\newcommand{\beq}{\begin{equation}}
\newcommand{\enq}{\end{equation}}
\newcommand{\beqa}{\begin{eqnarray}}
\newcommand{\beqast}{\begin{eqnarray*}}
\newcommand{\enqa}{\end{eqnarray}}
\newcommand{\enqast}{\end{eqnarray*}}

\def\GeV{\nobreak\,\mbox{GeV}}

\begin{document}

\begin{frontmatter}



\title{Energy dependence of  proton-proton elastic scattering at  large momentum transfer}


\author[first]{Erasmo Ferreira}
\affiliation[first]{organization={Instituto de Física, Universidade Federal do Rio de Janeiro},
            addressline={C.P. 68528}, 
            city={Rio de Janeiro},
            postcode={21945-970}, 
            state={Rio de Janeiro},
            country={Brazil}}

\author[second]{Anderson Kendi Kohara}
\affiliation[second]{organization={ AGH University Of Science and Technology, Faculty of Physics and Applied Computer Science},
            addressline={Mickiewicza 30}, 
            city={Kraków},
            postcode={30-059}, 
            country={Poland},
          }

\begin{abstract}

The measurements of proton-proton elastic scattering for large momentum transfer 
at energies in the range $\approx$  20 to 60 GeV  show a simple behaviour of 
 form $d \sigma/dt \approx  {\rm const}~|t|^{-8}$, 
 apparently with  no energy dependence. 
 In the present work detailed analysis of the data  shows a decrease  of the magnitude 
of the tail with the energy, still with preservation of the power $|t|^{-8}$. 
The analysis allows the definition of a    band 
for the energy dependence of the amplitude  with central line $ A(\sqrt{s}) = 0.389 \log{(100)}/\log{\sqrt{s}} ~ \GeV^6 $ .  
The rate of decrease  describes   well the data 
at the   distant energy $\sqrt{s}$ = 13 TeV, with reduction of the cross section 
by a factor 5.71. This result gives   prediction for new experiments 
at high energies, and opens  important question for theoretical
investigation. 
\end{abstract}




\end{frontmatter}




 \section{Introduction\label{intro}}

With basis on the data of Fermilab \cite{Conetti:1978cp} and ISR experiments
\cite{Nagy:1978iw,Amaldi:1979kd}, it  is usually believed that the differential cross section
of elastic pp and p\=p scattering for large momentum transfer 
has a form $|t|^{-8}$,  with no  energy dependence. 
The interpretation given for the $|t|^{-8}$ form is the dominance of 
perturbative process of three-gluon exchange \cite{Donnachie:1979yu}, while the magnitude has no simple interpretation, as   
the studied multiple exchanges predict   different behaviour, both 
in $\sqrt{s}$  and $t$ dependence.   
 
In the present work we treat the pp elastic data for large $|t|$ 
 in detail, and arrive at a slow energy dependence of the real amplitude, 
like $1/\log{\sqrt{s}}$, still preserving the $|t|^{-8}$ behaviour. 
The variation of the cross section in 
the rather small energy range, about  20 GeV to 60 GeV,  of the observed data 
has not received attention, but the  influence is very strong at the distant TeV energies.
We show that the prediction for  13 TeV , with a reduction by a  factor  2.39 
in the amplitude and a factor 5.71 in the cross  section, is in  good agreement 
with the data.  
It is observed   that the energy dependence  of the amplitude 
for large $|t|$ scattering   can be described by a power of the running strong coupling with scale $\sqrt{s}$ ,      
$\alpha_S(\sqrt{s})$. This information may be  useful for studies of the operating  mechanisms  
in pp elastic scattering  for large momentum transfer. 

The analysis of data, with analytical description of  the energy dependence 
of the amplitude, 
and confirmation of prediction for 13 TeV, are all presented in Sec. \ref{tail}.

In Sec. \ref{remarks} we comment on the results and on expectations. 
 
 \section{Energy dependence of the perturbative tail \label{tail}}

The first observation of similarity of pp scattering for large $|t|$ at 
different energies was  made in the comparison of data at $\sqrt{s}$ 
= 19.6 and 27.4 GeV in Fermilab \cite{Conetti:1978cp}. 
Measurements of the differential cross sections in  ISR/CERN at  
energies from 23.5 to 62.5 GeV   \cite{Nagy:1978iw}  indicated the existence of 
a regularity in the large $|t|$ range, with same form $ {\rm const} ~  |t|^{-8}$ and 
same magnitude   for all energies. 

The analysis of the ISR measurements was extended with higher accuracy \cite{Amaldi:1979kd}, 
and   a Fermilab experiment   at $\sqrt{s}$ = 27.4 GeV  was made in   very large  
momentum transfers with 39 points in the 
 range $ 5.5 \leq |t| \leq 14.2 ~ \GeV^2$   \cite{Faissler:1980fk},
confirming compatibility with the $|t|^{-8}$ behaviour, all  apparently with the 
same magnitude. The data are shown in Table \ref{table_tail}.    
 
On the theoretical side,  explanation for the $1/|t|^8$ behaviour of $d\sigma/dt$ 
in terms of a real three-gluon exchange  amplitude  was  soon given by Donnachie 
and Landshoff \cite{Donnachie:1979yu}. The diagram with the three quarks of each proton exchanging one gluon, assuming that each of these individual 
exchanges is made with the same fraction of the transferred momentum, leads to 
\begin{equation} 
\frac{d\sigma}{dt} \sim  \frac{[\alpha_S(t/9)]^6}{t^8} ~ , 
\label{three_gluon}
\end{equation}  
where  $\alpha_S$ is  the QCD strong coupling. 
The insertion of the diagram 
 in a model of multi Pomeron and Reggeon exchanges
was studied by Donnachie and Landshoff \cite{Donnachie:1979yu,Donnachie:1983hf,Donnachie:1996rq}~, 
and also other  mechanisms were analysed, such as triple singlet exchanges,
but the energy independence is not consistently explained. 
 Also the  $|t|$| dependence expected from the coupling  factor $[\alpha_S(t/9)]^6$ is not visible in the data. 
 The  phenomenology of pp scattering at large momentum transfer  is 
 a complicated and open question, involving QCD complexities, 
 including the proton structure.

In  general, proton-proton elastic  scattering for all $|t|$ is described  with a 
non-perturbative complex amplitude $T^{\rm NP}$ covering the forward peak  
and the dip-bump structure in the cross section. The large $|t|$  
range is described with an additional    real term, that 
dominates the real and imaginary parts of  $T^{\rm NP}$ for large $|t|$. 
 Although there are inconsistencies and may be there are other mechanisms, 
 we define in the present work an amplitude term $R_{ggg}\sim |t|^{-4} $,  called {\it three-gluon exchange} 
 to follow the original suggestion of a simplified three-gluon exchange   mechanism.   

  Characteristic features of the $R_{ggg}$  
  amplitude are  that 
it is real, has opposite signs for pp and p\=p ~  scattering,
and the form $1/|t|^4$. 
Other possibilities of exchanges exist \cite{Donnachie:1983hf,Donnachie:1996rq}, but the specific property 
of the change of sign for p\=p scattering 
seems to be fulfilled, according to the experiment at 53 GeV \cite{Breakstone:1985pe} 
where both pp and p\=p  scattering are measured in the dip-bump range, that is
sensitive to the sign of $R_{ggg}$. 

With a separate real term of perturbative nature, called 
$R_{ggg}$, added  to the  complex  amplitude  of nonperturbative dynamics $T^{NP}$, 
we write for the amplitude of elastic pp and p\=p ~  elastic scattering   
\begin{equation}
T(s,t)=T^{NP}(s,t)+ R_{ggg}(s,t) ~.  
\label{amplitude}  
\end{equation}
The normalization  used is 
\begin{equation}
\frac{d\sigma(s,t)}{dt} =  (\hslash c)^2 T^2(s,t) ~  , ~ {\rm with} ~  T(s,t) ~ ~ {\rm  in} ~ ~ \GeV^{-2}   ~ . 
\label{norm}
\end{equation}
The nonperturbative part is given by the chosen model and includes the Coulomb 
interaction. It is assumed that, as occurs in several models, for large $|t|$ the 
perturbative real term is dominant over the nonperturbative  amplitudes so that  
it can be identified and studied directly with the data. 
 On the other hand, when  $d\sigma/dt$ is described for all  $|t|$,
 the uncontrolled increase of $|t|^{-4}$ in $R_{ggg}$  for small $|t|$  must 
be accompanied  with a  damping  factor. In the present work we are concerned  
only with large $|t|$ data, and a damping factor  is not  directly  used. 

 We wish to investigate a possible energy dependence of the $|t|^{-8}$ 
tail in  pp elastic scattering, and then we introduce a quantity $A(\sqrt{s})$ to be determined, and write  
   \begin{equation}
   R_{ggg}(t)\equiv\pm \frac{A(\sqrt{s})}{t^{4}} ~ , 
   \label{R_tail} 
  \end{equation}  
where the sign is minus for p\=p scattering.
The quantity  $A(\sqrt{s})$ has units  $\GeV^6$, and $t$ has units $\GeV^2$.
   The  differential cross section in the large $|t|$ domain is written
\begin{equation}
\frac{d\sigma(s,t)}{dt} \simeq  (\hslash c)^2 R^2_{ggg}(s,t)  ~~ , ~~~ {\rm large} ~ |t| ~ . 
\label{cross}
\end{equation}
With   $R_{ggg}(s,t)$ in $\GeV^{-2}$ units  and $d\sigma/dt$ in mb/$\GeV^2$,  ~   
  we have $(\hslash c)^2~ =  ~ 0.389379 ~{\rm{mb}} ~ \GeV^2 $ ~ .  
The form $|t|^{-4}$ 
written  for the $|t|$ dependence in Eq.(\ref{R_tail}) is actually confirmed 
as very good at each energy in the interval 23.5 to 62.5 GeV
examined. This is a partial support for the original simple form assumed
for the mechanism of three particle exchanges in Eq. (\ref{three_gluon}), with absence of the factor with strong coupling, that is not required by the data.   
 
 Through a detailed examination of data
in the energy  range  23.5 GeV to 62.5 GeV where measurements  
at large $|t|$ were made \cite{Amaldi:1979kd,Faissler:1980fk}, we show  that 
the traditionally assumed  energy independence for large $|t|$ is an approximation, 
while  actually the magnitude of the  tail decreases slowly with the
scattering energy. We obtain a simple  form for   $A(\sqrt{s})$ with a 
             $(\log{\sqrt{s}})^{-1}$ dependence.  

 The data used  in our analysis are shown in 
Table \ref{table_tail} with the fitted values of $A(\sqrt{s})$.  
In Fig. \ref{tail_energy_dep} we plot the values of  $A(\sqrt{s})$ 
for the six energies investigated.  
In the plot we identify a band limited by regular 
tendencies in the  decreasing  values. We obtain a parameterization for the central line   
\begin{equation} 
A(\sqrt{s})= 0.389 \frac{\log{(100)}}{\log{\sqrt{s}}} ~, 
\label{energy_param}  
\end{equation}
with $A(\sqrt{s})$ in $\GeV^6$ and $\sqrt{s}$ ~ in $\GeV$ .
The dashed higher and lower lines defining the uncertainty band are  respectively 
\begin{eqnarray}
 A_{\rm plus}(\sqrt{s}) &=& 0.460 ~ \frac{\log{(52.8)}}{\log{\sqrt{s}}} \label{band_up} \\
{\rm and }                                                        \nonumber          \\ 
 A_{\rm minus}(\sqrt{s}) &=& 0.4246 ~ \frac{\log{(62.5)}}{\log{\sqrt{s}}}   ~ .  
 \label{band_down}
\end{eqnarray}

The log forms give small changes in the range of the data in Fig. \ref{tail_energy_dep}, 
but  gives remarkable differences  for TeV  energies. For example, for 13 TeV   
we  obtain $A(\sqrt{s})= 0.1890 \pm 0.0037$ $\GeV^6$ , that is  about $\log(13000)/\log (52.8)$= 2.39 
times smaller than  the value 0.460 (see 52.8 GeV in the Table). For the cross section the 
reduction is by a factor 5.71 ~ . 
 
Fig. \ref{tail_13_TeV-fig}  shows the large $|t|$ data at 13 TeV \cite{TOTEM:2018hki}
described in solid line with the new prediction of Eq.({\ref{energy_param}). 
In the second part of the same figure we compare the data for 52.8 GeV and 13 TeV showing 
the strong reduction in the cross sections.
The attempt of direct  connection of 13 TeV data with points of Faissler et al \cite{Faissler:1980fk} 
at 27.4 GeV, thus without energy dependence,  was highly unsatisfactory  \cite{Ferreira:2020nyi} . 

We have a formidable prediction for the energy dependence of  pp elastic scattering 
for large momentum transfer, covering with  good accuracy 
the energy scale with a  factor of more than 200, namely from 62.5  GeV to 13000 GeV. 

It is interesting to relate the magnitude in Eq.(\ref{energy_param}) with the QCD strong coupling. 
Fig. \ref{scale_fig} shows the values of  $\alpha_S(\sqrt{s})$ from  the Particle Data Group \cite{Workman:2022ynf}
in the interval from 10 to 1000 GeV. With $\sqrt{s}$ for scale,
the values are parameterized with good accuracy in the form 
\begin{equation}
\alpha_S(\sqrt{s})= 0.1179~ \Big[ \frac{\log{(91.79)}}{\log{\sqrt{s}}}  \Big]^{0.637}   ~ . 
\label{alpha_param}  
\end{equation} 
Notice that at $m(Z_0)=91.79 ~ \GeV $, $\alpha(m(Z_0))=0.1179$ has its best determination. 
This is a very convenient parameterization for $\alpha_S(\sqrt{s})$, as we obtain simply 
\begin{equation}
A(\sqrt{s})= 11.386 ~ [\alpha_S(\sqrt{s})]^{1.57}  ~  ,
\label{A-alpha}   
\end{equation}
with $A(\sqrt{s})$  in $\GeV^6$ ~ and $\sqrt{s}$ in $\GeV$.
This is a compact and precise phenomenological result of the present work. Thus we say : 
the energy dependence of the three-gluon exchange amplitude 
$|t|^{-4}$ goes with a simple 
power 1.57 of $\alpha_S(\sqrt{s})$. We cannot tell whether this a
simple curiosity or a useful inspiration in the search for a dynamical 
mechanism.  
\begin{figure}[b]
        \includegraphics[width=8cm]{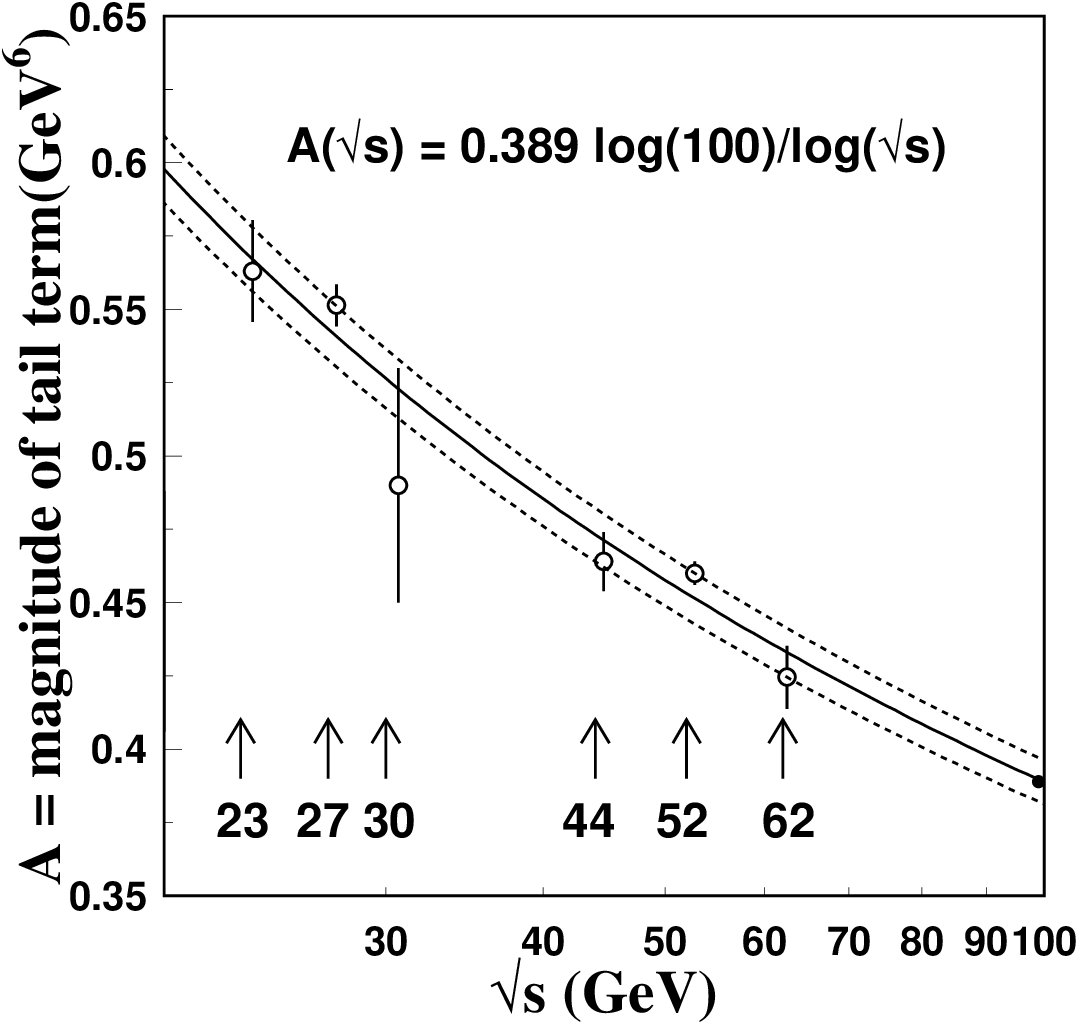} 
        \includegraphics[width=8cm]{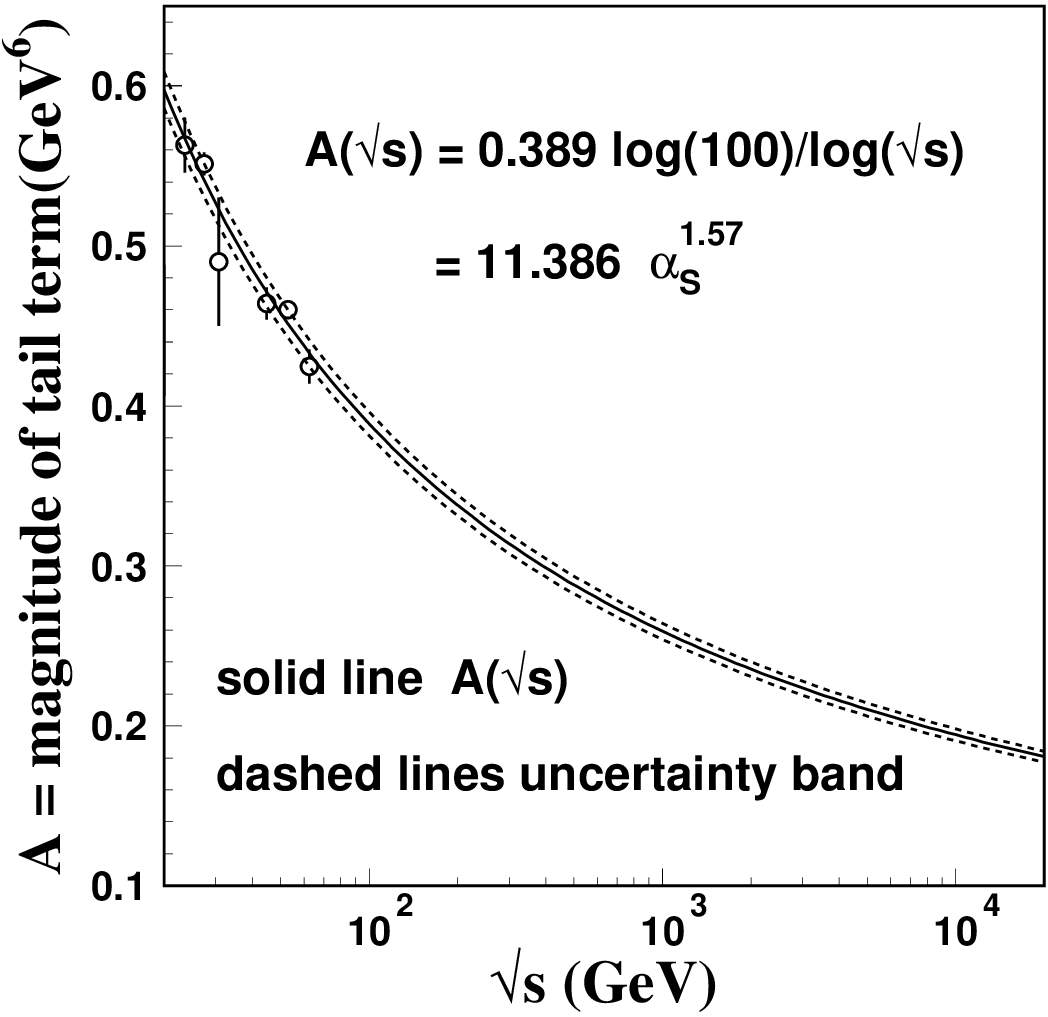} 
 \caption{ Energy dependence of the  tail term  
$A(\sqrt{s})/t^4$,   of the large $|t|$  elastic pp  amplitude, as is Eq. (\ref {energy_param}).        
The dashed lines show the uncertainty  band defined in Eqs.(\ref{band_up},\ref{band_down}). The marked points 
represent the values of $A(\sqrt{s})$   at each energy, according to Table \ref{table_tail}  .
The second plot shows the extension for higher energies, with  prediction of the value $A(\sqrt{s}) = 0.189 \GeV^6$ at 
13 TeV, confirmed with the data in Fig. \ref{tail_13_TeV-fig}.     
    }
\label{tail_energy_dep}
\end{figure}

\begin{figure}[b]
  \includegraphics[width=8cm]{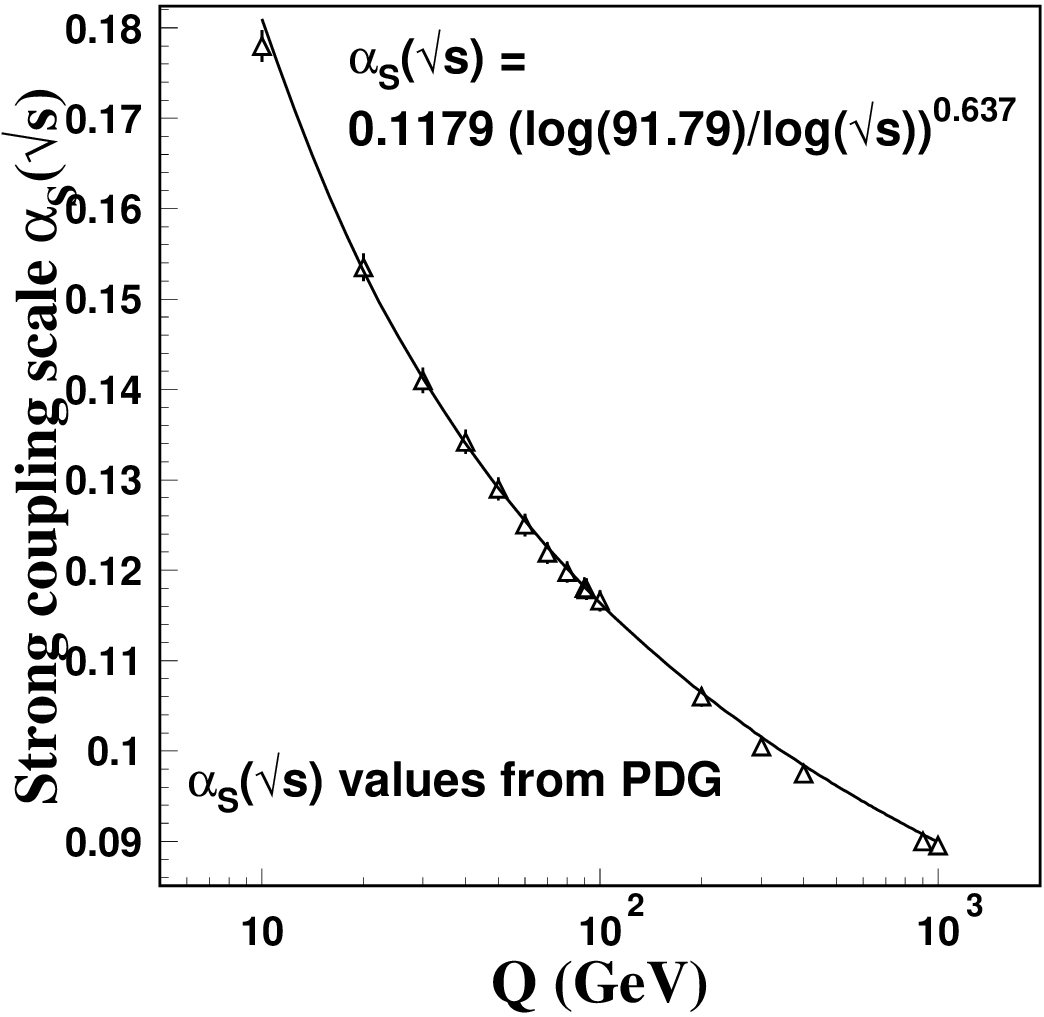} 

\caption{Parameterization in Eq. (\ref{A-alpha}) of the scale dependence of the strong coupling, using 
analytical form similar to the   the energy dependence of  $A(\sqrt{s})$.
The $\alpha_S(\sqrt{s})$ values are from Particle Data Group \cite{Workman:2022ynf} using $\sqrt{s}$ for the scale. 
 Notice that at $m(Z_0)=91.79 \GeV$, $\alpha(m_{Z_0})=0.1179$ has its best determination, used as reference value for the parameterization. The error bar of the exponent is 
 $0.637\pm 0.009$, and the fit has $\chi^2=0.4407$ ~ . 
     }
 \label{scale_fig}
\end{figure}

{\small
\begin{table*}[ptb]
\caption{ $\chi^2$ (namely $\chi^2/d.o.f.)$ values for the fittings of points of 
large  $|t|$ 
in the measurements of pp elastic scattering  in ISR at energies 23.5, ~ 30.7 , ~  44.7, ~ 52.8  
and ~ 62.5   GeV  \cite{Amaldi:1979kd} and for 27.439  GeV from Fermilab \cite{Faissler:1980fk},  
  with  form $ R_{ggg}= A(\sqrt{s}) |t|^{-4}$ in the amplitude. This form, called "tail"  
is based on  the idea  of 3-gluon exchange  perturbative contribution \cite{Donnachie:1979yu}   
to be  added to the   basis given by a nonperturbative dynamical model.  Only statistical errors  
in the reported experimental data are considered in the calculation  
of $\chi^2$ values shown  in the table.}
       \begin{tabular*}{\textwidth}{@{\extracolsep{\fill}}cccccc@{}}
 $\sqrt{s}$ & $|t|$ range & N      & $A$           &$\langle\chi^2\rangle$(stat) & References  \\ 
 $\GeV$     & $ \GeV^{2}$ & points &$\GeV^{6}$     &                             &     \\ \hline 
 23.5      & 3.45 - 5.75 &  17  &$0.563\pm 0.030$& 0.565               & \cite{Amaldi:1979kd}  \\  
  27.4      & 5.5 - 14.2  &  39    &$0.551 \pm 0.007$ &   1.871         & \cite{Faissler:1980fk}   \\  
   30.7     & 3.55 - 5.75 &  17  &$0.490\pm 0.040$  &     1.265        & \cite{Amaldi:1979kd}    \\  
  44.7      &  3.25 - 7.25  &  22 &$0.464\pm0.010$ &   1.476            &  \cite{Amaldi:1979kd}   \\  
   52.8     & 2.85 - 9.75  &  31 &$0.460\pm0.004$  &   5.734             & \cite{Amaldi:1979kd}    \\  
  62.5      & 3.05 - 6.25  &  22 &$0.425\pm0.011$ & 1.225               & \cite{Amaldi:1979kd}    \\  
  \hline 
\label{table_tail}   
 \end{tabular*}
\end{table*}
 }

\begin{figure}[b]
  \includegraphics[width=8cm]{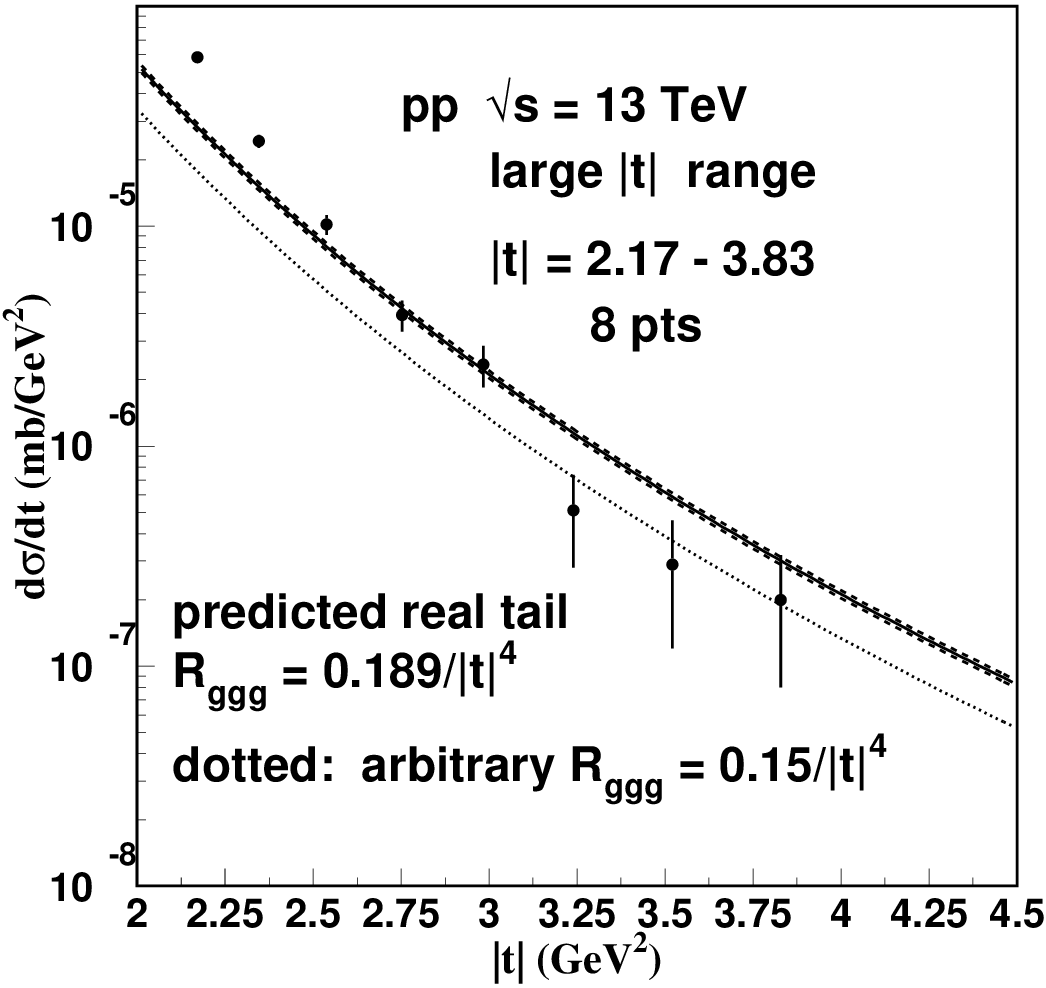} 
   \includegraphics[width=8cm]{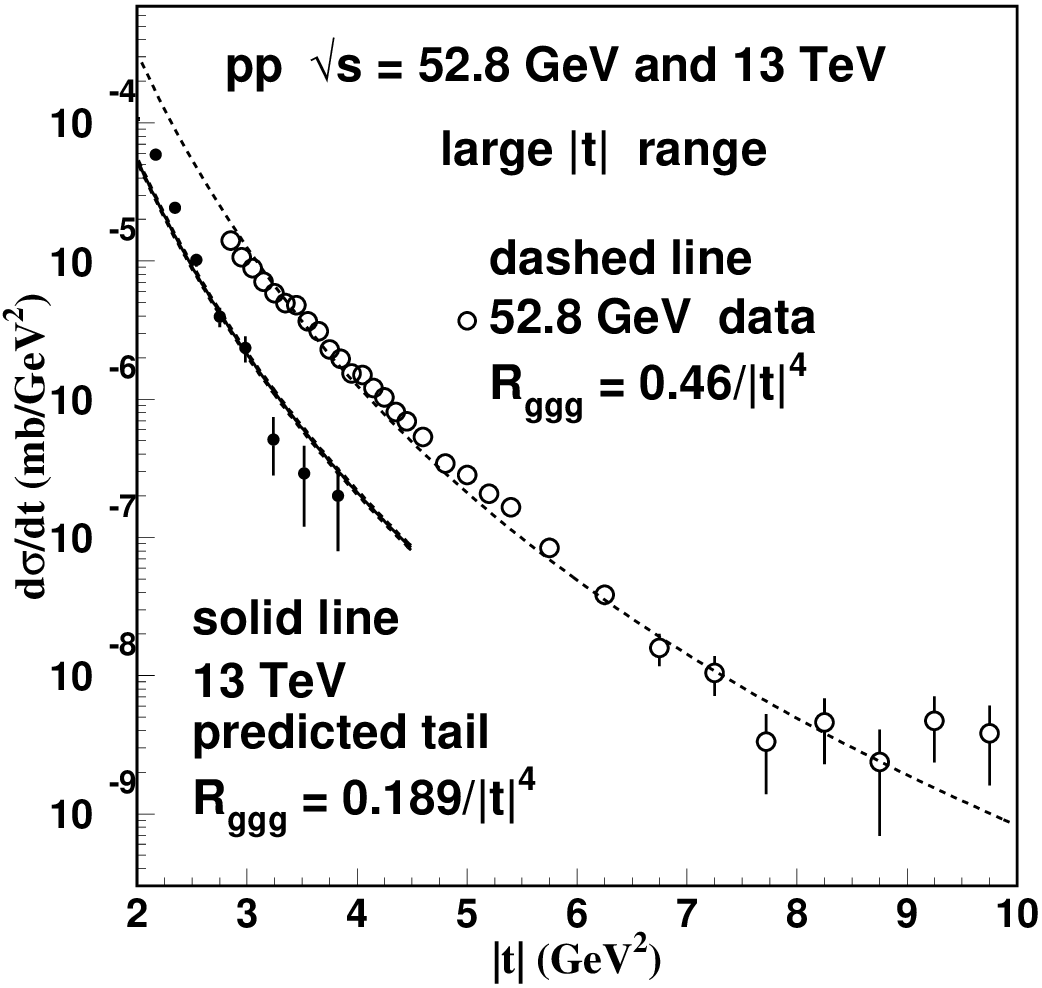} 
      \caption{ Prediction and data   
  \cite{TOTEM:2018hki} for large $|t|$ elastic scattering at 13 TeV. The thick solid line represents 
   $d\sigma/dt = 0.3894 ~ R^2_{ggg} ~ {\rm mb}/\GeV^2$ , 
with the value $A(\sqrt{s})=0.1890\pm 0.0037 ~ \GeV^6$ including the uncertainty band.   
  To provide a sense of the influence of the predicted value  for $A$, we show the  first plot with  
 dotted line the   curve with $A(s)=0.15 ~ \GeV^6 $.
We remark that these curves are calculated with the term  $R_{ggg}$ alone, with no 
  contributions that would come from a model describing small 
and mid $|t|$ ranges and that are  small  for  large $|t|$. 
 These contributions  may have influence in the first three 
data points ($|t|\leq 2.5 \GeV^2$) in the lower end of the shown range.
In the second plot we put together the 52.8 GeV data (with $A=0.460$)  and  13 TeV   
 data (with $A=0.189$)   
to compare the magnitudes of the respective cross sections,
illustrating clearly the energy dependence of the large $|t|$ tail.  The few points of the 13 TeV data and their error bars 
do not allow a precise confirmation of the $1/|t|^8$ behaviour 
in the tail. 
} 
\label{tail_13_TeV-fig}
\end{figure}

 \section{Remarks} \label{remarks}

The regularity of the form $1/|t|^8$ in the differential cross section in 
pp elastic scattering  cross section for large momentum transfer about $|t|\geq 3 \GeV^2$
for the energies $ 23.5 \leq \sqrt{s} \leq 62.5$ is recognized since 1979 
\cite{Conetti:1978cp,Nagy:1978iw,Donnachie:1979yu}.    
In this rather small energy interval the tails of the $|t|$ distributions, 
put together in a plot of $d\sigma/dt$, seems to be in a unique narrow band. Thus 
Fig.1 of the article of Donnachie and Landshoff in 1996 \cite{Donnachie:1996rq} shows  
all data for all ISR  energies running 
almost together, with the unified description $d\sigma/dt=const/|t|^8$ .
This form  suggested  \cite{Donnachie:1979yu}  a mechanism of exchange of three gluons or alternatively of three singlets, 
 but the energy dependence
 and the absence of the influence of the $\alpha_S(|t|/9)$ couplings 
 in Eq. (\ref{three_gluon}) are not explained.  
 
We split the band of large $|t|$ data separating the energies, 
and show  that the constant is an approximation. 
As the form $|t|^{-8}$ is well preserved at all energies, we write Eq.(\ref{R_tail}),
and obtain  Eq.(\ref{energy_param}) 
from the numerical values in  Table \ref{table_tail} .
The decrease with the energy  is slow enough to have been 
overlooked in the  small   range of the ISR data, but the reduction in the 
cross section from 52.8 GeV of ISR to 13 TeV of LHC is of a factor 
$ [\log(13000)/\log(52.8)]^2=5.71  ~  $
according to Eq. (\ref{energy_param}). This decrease is remarkably confirmed in 
Fig. \ref{tail_13_TeV-fig}  with the plot of the data of these two energies. 

It is  interesting  that the parameterized form of $\alpha_S(\sqrt{s}) $  
in Eq.(\ref{alpha_param})  
leads to the form of Eq.(\ref{A-alpha}) for the energy dependence    
in terms of the running strong coupling. Hopefully this relation may 
inspire investigations of the underlying QCD mechanisms.

\subsection{ Damping factor and influence on the dip-bump region . \label{damping} } 
 
In calculations of $d\sigma/dt$ for all $|t|$ the term $R_{ggg}$ 
in Eq. (\ref{amplitude}) must be accompanied by a damping factor 
to avoid the uncontrolled increase for not-large $|t|$ . 
We  suggest a form 
\begin{equation}
R_{ggg}(s,t) = \pm \frac{A(\sqrt{s})}{|t|^4 } [1-e^{-0.005 ~ |t|^4}][1 -e^{-|t|}] ~ .   
\label{damp}
\end{equation}
At ISR energies the dip  in $d\sigma/dt$ is located  at $|t|\approx 1.3 ~\GeV^2$, thus 
not very distant from the range where $R_{ggg}$ is more influential. In the dip, 
both   imaginary and real parts of $T^{NP}(s,t)$ are
small, and  $R_{ggg}$, with appropriate damping factor, has influence on the form of the dip-bump structure.

 A characteristic feature of $R_{ggg}$ is the opposite signs in pp and p\=p ~ scattering. The non-perturbative 
basis is nearly  identical for the pp and p\=p ~ systems, so that the main difference between the two systems is due to the 
 sign of  the term $R_{ggg}$. The dip in $d\sigma/dt$ is formed 
by the relative positions  of the zeros of the  real and imaginary parts of the $T^{\rm NP}$ amplitude. 
The effect of the  $\pm$ sign  , causing  
the displacement of the zero of $Re[T^{\rm NP}]$, 
is a flattening of the dip in p\=p ~scattering \cite{Pereira:1999ba}. 
This change is observed at 53 GeV \cite{Breakstone:1985pe} where both  pp and 
p\=p ~ are measured  in the   dip-bump region of $d\sigma/dt$. 
 
In known models, the real part is positive for large $|t|$, and in pp the
additional positive  term for the tail interferes constructively with it,
putting  the real zero closer to the imaginary zero. In p\=p it is 
the contrary. Thus the term $R_{ggg}$ has sign +1 for pp and 
sign -1 for p\=p ~. 

At   TeV energies the dip in $d\sigma/dt$ 
is much higher and far away, at $|t| \approx 0.5 ~\GeV^2$, and is not 
influenced by the damped $R_{ggg}$ of Eq.(\ref{damp}). 
The pp amplitudes at  TeV  energies  fall very rapidly, 
and for large $|t|$ around  3 to 4 ~  $ \GeV^2 $  has values around
those from $R_{ggg}$, so that  the three-gluon exchange 
determines values at the data. We show this for 13 TeV  in 
Fig. \ref{tail_13_TeV-fig} .

\subsection {The need of a specific tail term. \label{others} } 

The treatment of the region of large $|t|$ in pp  elastic scattering requires 
a specific additive complement to produce the observed tail. 
The first example is the model of multi-exchanges by Donnachie and Landshoff \cite{Donnachie:1996rq,Donnachie:2013xia} who first introduced the term with three-gluon exchange in the amplitude.
  
In  the  work by A.A. Godizov \cite{Godizov:2021ksd}, the calculation in a Regge-eikonal formalism, the calculations with  a Soft Pomeron   
and a secondary Reggeon does not fit well the large $|t|$ data at the ISR energies, as shown in Fig. 3 of this paper.  A regular difference 
with respect to data exists for all the  energies  from  23.4 GeV to 62.5 GeV  
 for $|t|\geq 2 \GeV^2$, apparently indicating that a term like the three-gluon exchange is missing.

Similarly in the work by O. V. Selyugin \cite{Selyugin:2012np}, the calculation for 52.8 GeV 
shown in Fig. 3 is below the data for $|t|\geq 3 \GeV^2$, also missing  a contribution 
like $|t|^{-8}$ . 

In the work by Gon\c calves and  Silva \cite{Goncalves:2018nsp} with the Phillips-Barger potential
 model the calculations are not appropriate for the  large $|t|$ range and  remain much below  the data. 

In a treatment  of 13 TeV Totem  data with L\'evy imaging method
\cite{Csorgo:2019egs} the evidence for a missing term to cover the large $|t|$ range is very neat. 
 
To understand   the role of the term $R_{ggg}$ in the description of $d\sigma/dt$ it is important that the  non-perturbative  model  provides the  complex amplitude disentangled  in its real and imaginary parts, with explicit  exhibition  of the amplitudes.

\subsection {Final Comments  \label{comments)} }  

 The observed energy and $|t|$  dependence in Eq.(\ref{energy_param}) or  Eq.(\ref{A-alpha})   
opens   questions for theoretical  investigation in terms of QCD structure and interactions. 
The   description of the universal $|t|^{-8}$  behaviour as 
a {\it three-gluon exchange}  mechanism in Eq. (\ref{R_tail}) 
omits the $ [\alpha_S(t/9)]^6 $   factor in Eq. (\ref{three_gluon}), that does not seem   visible 
in our phenomenology. The reason for the weakness of the $|t|$ dependence of this factor is an important problem \cite{Donnachie:1996rq}.

For high energies the predictions  suggest  measurements in a delicate region where cross sections 
have very small values. The data  of pp elastic scattering for large $|t|$ are very scarce, 
restricted to the 8 points  in Fig. \ref{tail_13_TeV-fig}. Few points and large error bars do not allow to
confirm surely whether the pure $1/|t|^8$ dependence in the tail  holds also at  large energies. It may  happen that we must write $A(\sqrt{s},t)$ in Eq. (\ref{R_tail})  and 
hopefully recover  Eq. (\ref{three_gluon}).  

Proton-proton elastic scattering is very important  in all range from very forward to large $|t|$, with important questions everywhere. Expectations for detailed and precise data   are  put in the RUN 3 of LHC.   

\section{Acknowledgements}
E.F.  wishes to thank the Brazilian agency CNPq   for financial support.   
Part of the present work was developed under the project INCT-FNA Proc. No. 464898/2014-5.
A.K.K. is supported  by the National Science Centre in Poland, grant no. 2020/37/K/ST2/02665. 
Part of the research leading to these results has received funding from the Norwegian 
Financial Mechanism 2014-2021.



\bibliographystyle{elsarticle-num}

\bibliography{references}






\end{document}